\documentclass[twocolumn, preprintnumbers, amsmath, aps, prl, showpacs]{revtex4}
\usepackage[dvips]{graphicx}
\usepackage{bm}
\usepackage{shapepar}

\begin{document}

\title{Zero-dimensional spin accumulation and spin dynamics in a mesoscopic metal island}

\author{M. Zaffalon}
\author{B.J. van Wees}
\affiliation{Department of Applied Physics and Materials Science Centre\\
University of Groningen, Nijenborgh 4, 9747 AG Groningen, the Netherlands}

\begin{abstract}
We have measured electron spin accumulation at 4.2~K and at room temperature in an aluminium island with all dimensions (400~nm~$\times$~400~nm~$\times$~30~nm) smaller than the spin relaxation length. For the first time, we obtain uniform spin accumulation in a four-terminal lateral device with a magnitude exceeding the ohmic resistance in the island. By controlling the magnetisation directions of the four magnetic electrodes that contact the island, we have performed a detailed study of the spin accumulation. Spin precession measurements confirm the uniformity of our system and provide an accurate method to extract the spin relaxation time.
\end{abstract}

\maketitle
What happens if we inject spin-polarised carriers into a small non-magnetic island? This is an outstanding question in the rapidly developing field of spintronics\cite{awschalom}. In metallic systems with submicrometer dimensions, still many (of the order of $10^7-10^8$) electron spins are involved. They will behave uniformly if the dimensions of the island are smaller than the spin relaxation length, $\lambda_{sf} = \sqrt{D\tau_{sf}}$, where $D$ is the diffusion constant and $\tau_{sf}$ the spin relaxation time. We will show that in this regime, the spin accumulation dynamics depends only on $\tau_{sf}$ and becomes independent of the transport properties such as $D$. Previous studies on electrical spin injection and detection have focussed on four-terminal devices larger than the spin relaxation length\cite{johnson85, jedema01, jedema02} or in two-terminal pillar structures\cite{katine}.

Here we report the study of the injection of spins in an aluminium island fabricated with \textit{all} dimensions smaller than the spin relaxation length $\lambda_{sf}$. The island is weakly coupled to the four cobalt leads by means of tunnel barriers\cite{filip}. We will show that to first order, the system is zero-dimensional with respect to the spin and the induced spin polarisation in the island is uniform. These conditions correspond to the regime $\tau_{dif\!f} < \tau_{sf} \ll \tau_{esc}$, where $\tau_{dif\!f}$ is the time for the electron to diffuse in the island and $\tau_{esc}$ the time to escape into the cobalt leads. In addition, there are two other (position-dependent) contributions to the chemical potential of the island, smaller than the spin accumulation, arising from the the charge current (ohmic resistance) and the spin current.

\begin{figure}
\includegraphics[scale = 1]{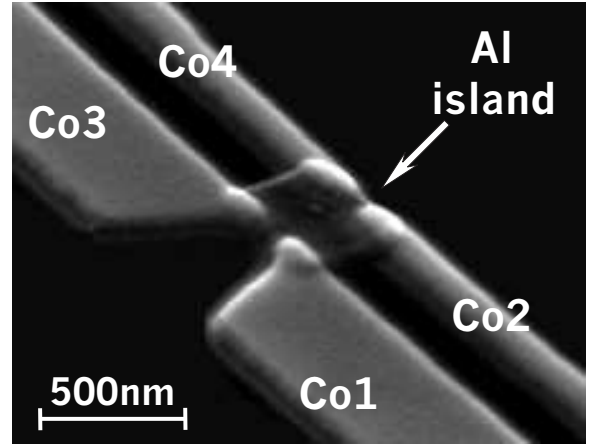}
\caption{\label{fig:sem} (a) Scanning Electron Microscope (SEM) picture of a device. The square aluminium island in the middle is contacted by four cobalt electrodes of different widths\cite{sem}.}
\end{figure}

In such a system, spin accumulation can be described in an elementary way and becomes the result of two competing processes: the injection of spins, and their dynamics and relaxation mechanisms. The injection into the aluminium island is obtained by driving a current in and out of two (e.g. electrode Co1 and Co2) of the four cobalt electrodes that contact the island, see Fig.~\ref{fig:sem}. Each current electrode \textit{i} carries a charge current $I$ and a spin current $\mathbf{I}_{m,i} = P I \mu_B\mathbf{m}_i/e$, where $\mathbf{m}_i$ is the magnetisation direction of the electrode. $P = (G^\uparrow - G^\downarrow)/(G^\uparrow + G^\downarrow)$ is the spin injection/detection efficiency of the tunnel barrier, with $G^\uparrow, G^\downarrow$ the tunnel conductances for the up and down spins, where up means oriented in the same direction as $\mathbf{m}_i$\cite{johnson85}. The spin relaxation mechanisms drive the out-of-equilibrium magnetisation inside the island back to equilibrium at a rate $\tau_{sf}^{-1}$.

The electrical detection of the spin imbalance is performed by using the two remaining cobalt electrodes, in this case Co3 and Co4. The signal detected by the electrode $i$ has a spin independent contribution $\mu_0(x) = \int\! f_0(\epsilon,x)\mathrm{d}\epsilon$ and a spin contribution $P\mathbf{m}_i \cdot \boldsymbol{\mu}$, with $\boldsymbol{\mu}=\int \!\mathbf{f}(\epsilon)\mathrm{d}\epsilon $, where $f_0$ and $\mathbf{f}$ are respectively the spin independent and spin dependent distribution functions\cite{brataas}. 

The spin accumulation contribution to the total signal in the 0D case depends on the injecting vector $\mathbf{m}_{inj} = \mathbf{m}_1 - \mathbf{m}_2$, the detecting vector $\mathbf{m}_{det} = \mathbf{m}_3 - \mathbf{m}_4$ and the volume of the island $\hat{V}$:

\begin{equation}
R_s = \frac{V}{I} = \frac{P^2\tau_{sf}}{e^2\nu_{DOS}\hat{V}}
\mathbf{m}_{inj} \cdot \mathbf{m}_{det}
\label{eq:spinaccumulation}
\end{equation}

where $\nu_{DOS} = 2.4 \cdot 10^{28}eV^{-1}m^{-3}$ is the aluminium density of states at the Fermi energy. For collinear electrodes, this reduces to the formula obtained by Johnson\cite{johnson93}. Spin accumulation occurs \textit{and} it is detected only if $\mathbf{m}_1$ and $\mathbf{m}_2$ are not parallel with each other and if $\mathbf{m}_3$ and $\mathbf{m}_4$ are also not parallel.

\begin{figure}
\begin{picture}(0, 0)
\put(0, 0){\normalsize a)}
\end{picture}
\mbox{\parbox{0.15\textwidth}{\includegraphics[width = 2.0cm, height = 2.7cm]{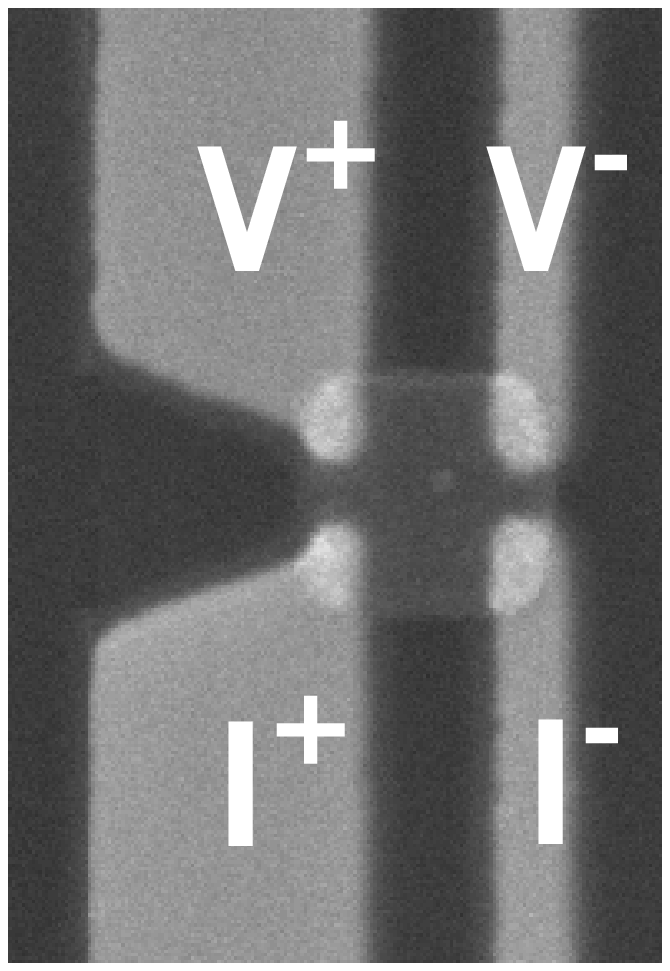}\\{\normalsize side}}
\hspace{0.0cm}
\parbox{0.15\textwidth}{\includegraphics[width = 2.0cm, height = 2.7cm]{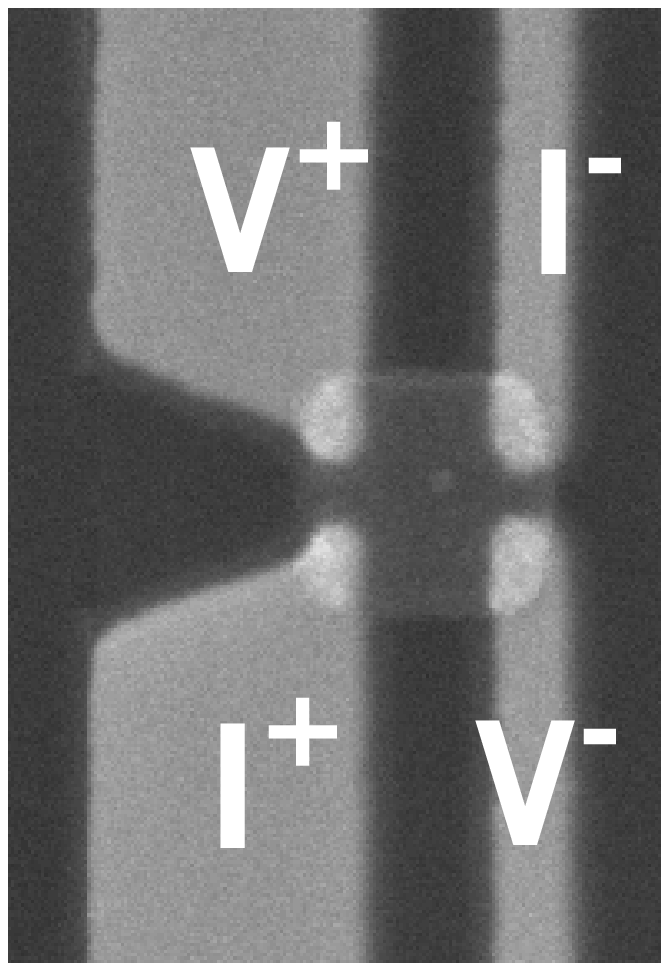}\\{\normalsize diagonal}}
\hspace{0.0cm}
\parbox{0.15\textwidth}{\includegraphics[width = 2.0cm, height = 2.7cm]{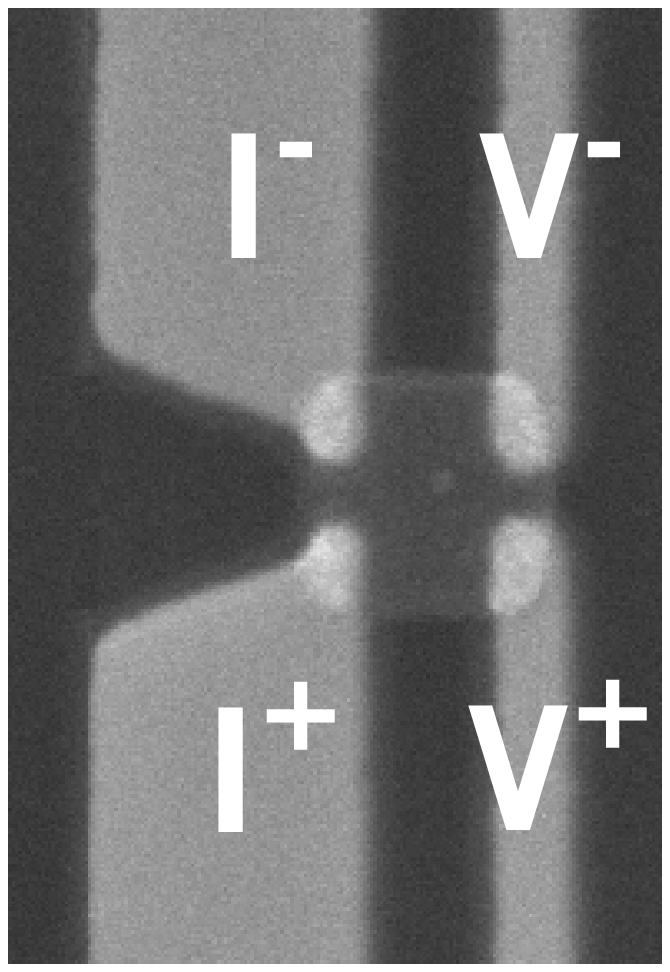}\\{\normalsize opposite}}}

{\vspace{0.3cm}}

\begin{picture}(0, 0)
\put(0, 0){\normalsize b)}
\end{picture}
\mbox{\parbox{0.15\textwidth}{\includegraphics[width = 2.0cm, height = 2.7cm]{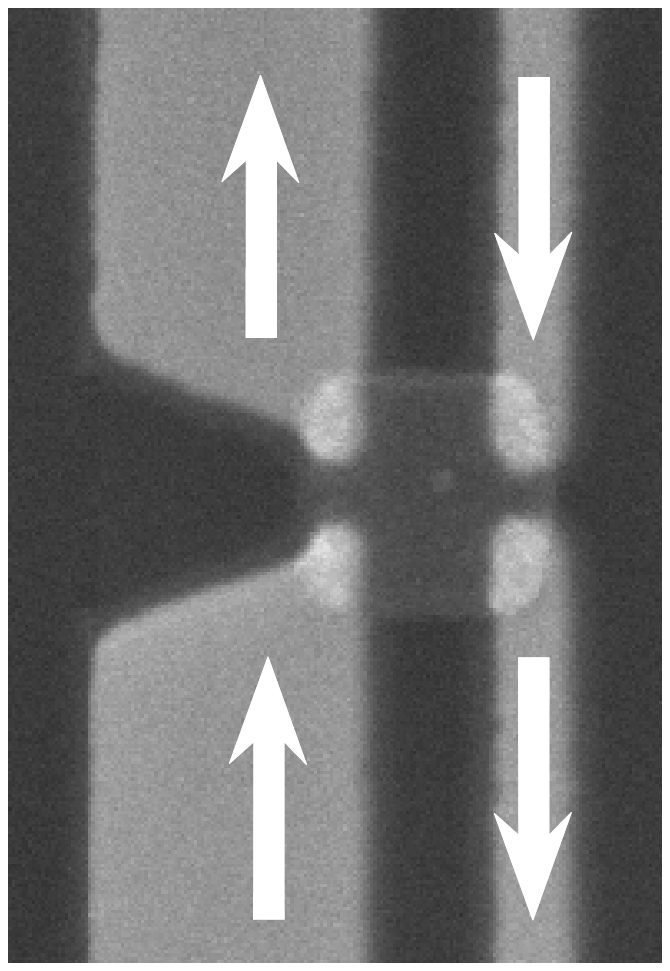}\\{\normalsize anti-parallel}}
\hspace{0.0cm}
\parbox{0.15\textwidth}{\includegraphics[width = 2.0cm, height = 2.7cm]{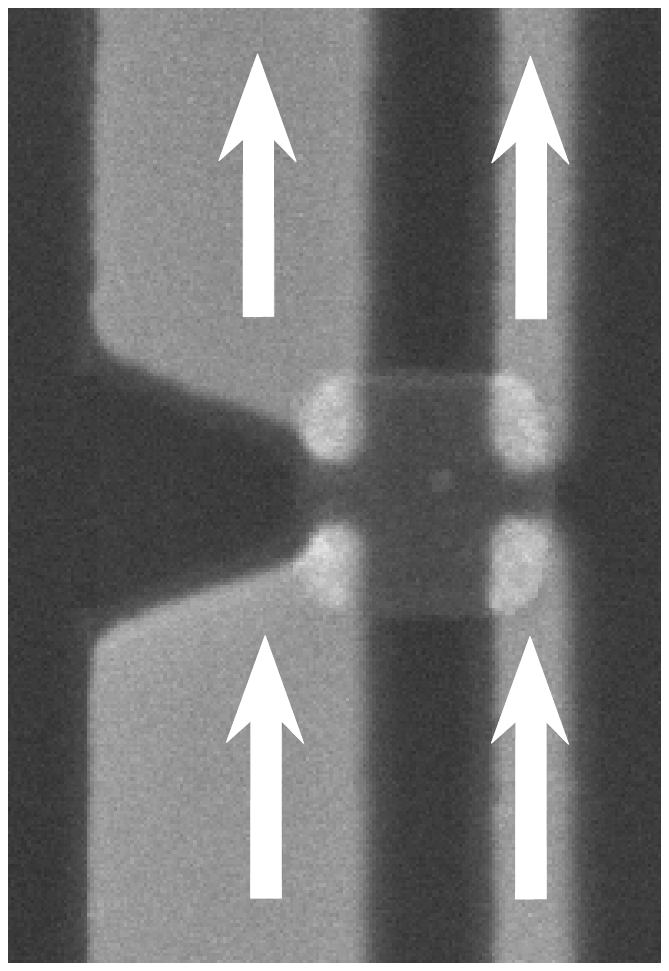}\\{\normalsize parallel}}
\hspace{0.0cm}
\parbox{0.15\textwidth}{\includegraphics[width = 2.0cm, height = 2.7cm]{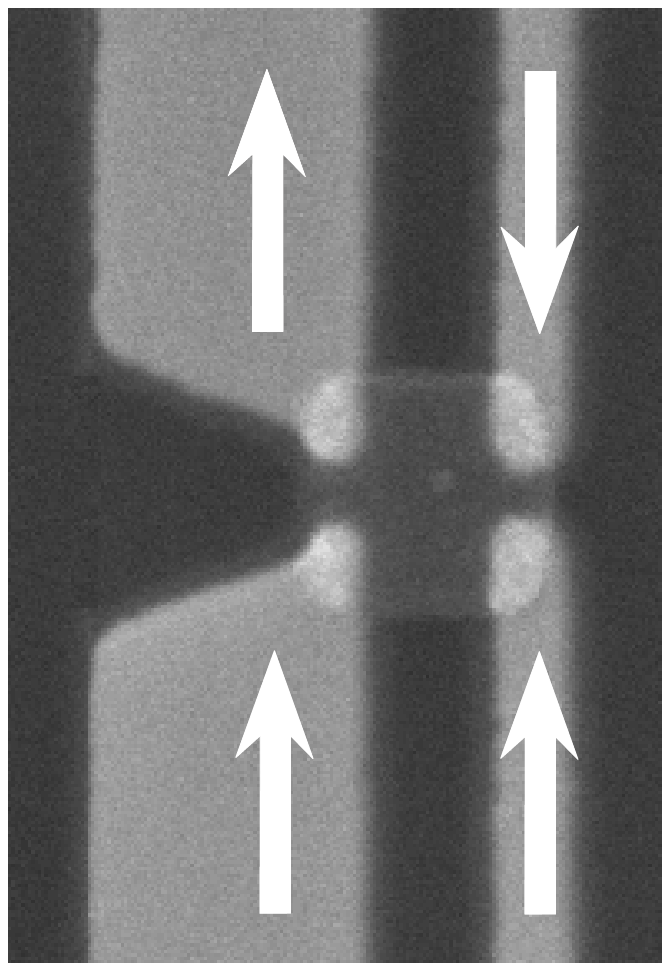}\\{\normalsize anomalous}}}

\vspace{0.3cm}

\begin{picture}(0, 0)
\put(-7, 40){\normalsize c)}
\end{picture}
\includegraphics[scale = 0.48]{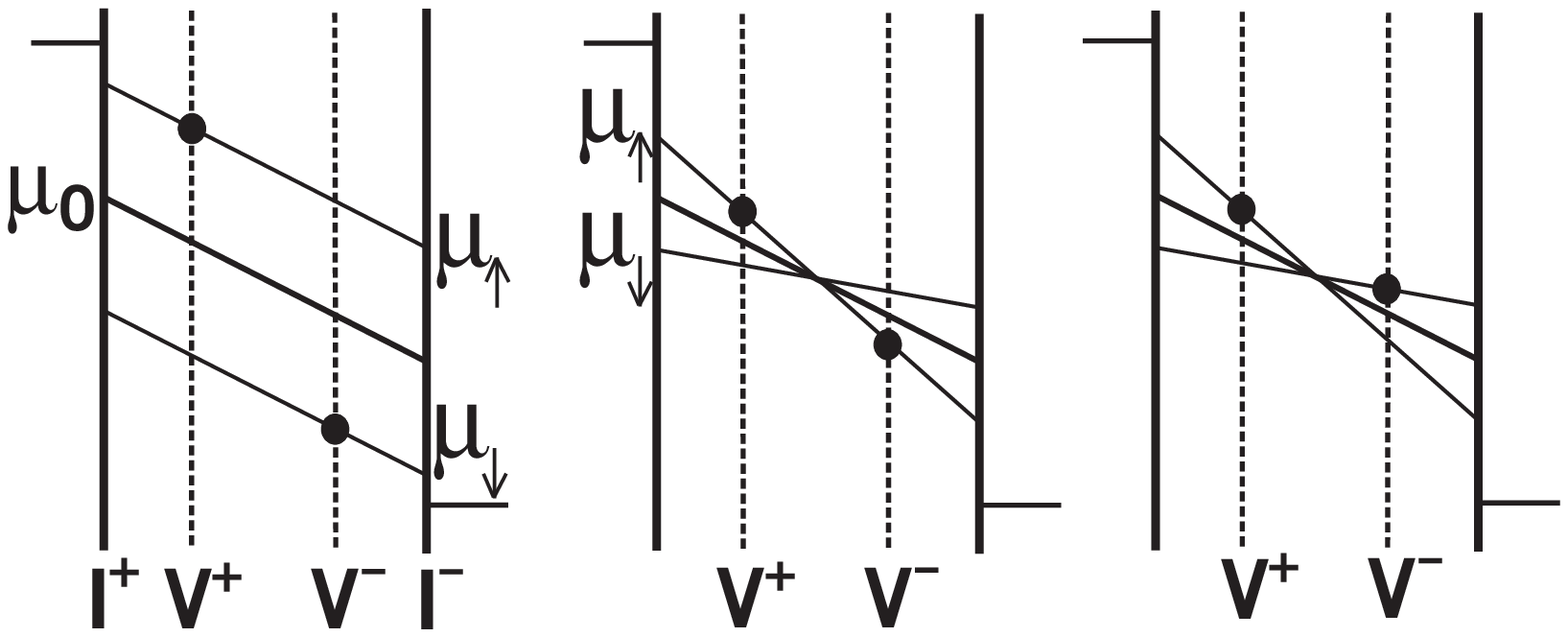}
\caption{\label{fig:chempot} (a) The three possible independent measuring configurations: current is sent from $I^+$ to $I^-$, the measured voltage $V$ is $V^+ - V^-$ . (b) The possible magnetic configurations and (c) the corresponding chemical potentials for the spin up and down inside the island for the ``side'' configuration: the thick line is the average chemical potential $\mu_0$, the two thin lines $\mu_{\uparrow, \downarrow}$. The black dots indicate the potential measured by the $V^+$ and $V^-$ probes for the case $P = 1$. In the parallel and anomalous configurations, an equal spin current density $|\mathbf{j}_m| = -\sigma_N \mu_B \cdot (\nabla \mu_\uparrow - \nabla \mu_\downarrow)/(2e)$ flows through the island.}
\end{figure}

The devices are defined by electron beam lithography and a two-angle shadow mask technique. A square aluminium island ($400\times400\times30$~nm$^3$) is deposited by \textit{e}-beam evaporation at a base pressure of $1 \cdot 10^{-6}$~mbar, through a suspended mask, followed by oxidization in pure oxygen ($2-10\times10^{-2}$~mbar for $1-5$~min) to produce tunnel barriers with resistances in the range $R_{TB} = 1-40$~k$\Omega$. Next four cobalt electrodes 40~nm thick are deposited from a different angle to contact the island. The electrodes have different widths, one pair 500~nm wide and one pair 100~nm, with the widest one having the lowest coercive field. Owing to the magnetic shape anisotropy, the electrode's magnetisation lies in the plane of the substrate, pointing in the positive or negative $\mathbf{\hat{y}}$ direction. By applying an in-plane external magnetic field, we can independently flip the magnetisation of the electrodes. We identify a ``parallel'' and an ``anti-parallel'' configuration, see Fig.~\ref{fig:chempot}(b). From the measurements, we conclude that a third, ``anomalous'' magnetic configuration also occurs, in which the two wide electrodes and one of the narrow ones are aligned, while the fourth one, probably due to a slightly different coercive field, is opposite.

The three possible electrical measuring configurations are depicted in Fig.~\ref{fig:chempot}(a): the current $I$ is sent from $I^+$ to $I^-$, the detected voltage is $V = V^+ - V^-$ and the signal we plot, $R = V / I$\cite{reciprocity}.

To describe spin transport and spin accumulation, we assume for the moment that magnetisation direction in the island is collinear with the electrodes, with $\uparrow$ directed in the positive $\mathbf{\hat{y}}$ direction. Then the island spin dependent chemical potentials are represented in terms of $\mu_{\uparrow,\downarrow}$\cite{valet, brataas, hernando}. Fig.~\ref{fig:chempot}(c) gives a schematic picture of the chemical potentials and the voltage contacts. In all three cases, the potential drop given by the charge current is $\Delta \mu_0 = eR_{ohm}I$, where $R_{ohm}$ is the island four-terminal ohmic resistance.

For the spin contribution, we analyse the three cases separately. In the anti-parallel configuration the spin accumulation signal is given by eq.~(\ref{eq:spinaccumulation}). In the parallel configuration, no net (average) spin accumulation occurs. However, spin-polarised carriers injected at Co1 and extracted at Co2 give rise to a the spin current of magnitude $|\mathbf{I}_m| = P I \mu_B/e$ that traverses the system, causing the spin polarisation to be space dependent. It can be shown that this gives a contribution $\delta R = P^2R_{ohm}$. This can be understood by considering that one of the two spin channels is partially used and the total conductance of the island decreases. In the limiting case of $P = 1$, the total conductance would halve. In the anomalous configuration, the above contribution to the resistance cancels, as can be seen in Fig.~\ref{fig:chempot}(c) and only the ohmic resistance is detected.

\begin{figure}
\includegraphics[width = 0.5\textwidth, height = 0.67\textwidth]{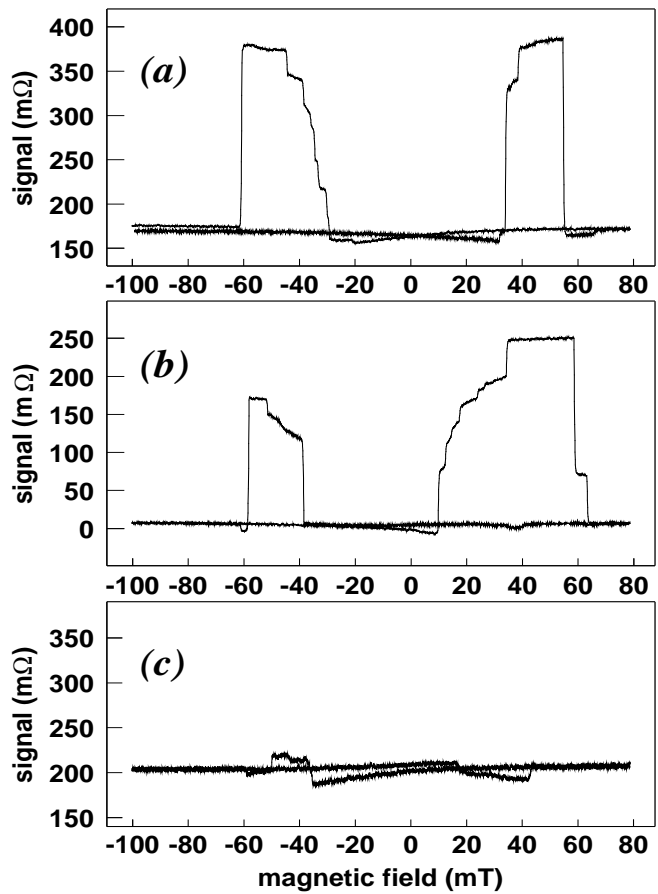}
\caption{\label{graph:meas4K} Measurements of $R = (V^+ - V^-) / I$ as a function of the in-plane magnetic field at 4.2~K, (a) in the ``side'', (b) ``diagonal'' and (c) ``opposite'' configuration.}
\end{figure}

The measurements were performed by standard low frequency lock-in techniques with a modulation current of $10-100~\mu$A. We have measured 6 devices in detail, of which one only at 4.2~K, two both at 4.2~K and at room temperature and three at RT only. For the last three we have also performed precession measurements (discussed later).

Fig.~\ref{graph:meas4K} shows measurements for the three configurations at 4.2~K in device A, with all tunnel barriers having a resistance of 20~k$\Omega$. The magnetic field is applied in the $\mathbf{\hat{y}}$ direction. We start with the field at $-100$~mT, so that the electrodes magnetisations are aligned in the negative $\mathbf{\hat{y}}$ direction. Ramping the field to positive values, we observe a sudden increase of the signal at $+30$~mT, when the magnetisation direction of the widest pair reverses. The magnetic configuration is now anti-parallel, spin accumulation occurs and the four terminal resistance is enhanced. When the second pair of electrodes also switches at $+60$~mT, the magnetisation configuration is again parallel but with all magnetisations pointing in the positive $\mathbf{\hat{y}}$ direction and the signal drops again. For the ``side'' configuration, the spin accumulation signal is 220~m$\Omega$ and it is slightly larger, 250~m$\Omega$ for the ``diagonal'' one.

The steps visible on the measurements at the switching of the larger electrodes are interpreted as the steplike reversal of the electrodes magnetisation, and therefore as discrete changes of the injector and detector vectors, $\mathbf{m}_{inj}$ and $\mathbf{m}_{det}$. In some cases, we observe imcomplete switching of the electrodes, see Fig.~\ref{graph:meas4K}(b), in which the left peak does not reach full height: this occurs mostly at 4.2~K.

Graph~\ref{graph:meas4K}(c) shows no spin signal for the ``opposite'' configuration: the two larger electrodes are always parallel as they flip at the same time. A similar behaviour was also observed for a device with tunnel barrier resistances between 15 and 35~k$\Omega$ and for a device with tunnel barriers of $3-5$~k$\Omega$.

These results are consistent with the assumption of an almost uniform spin accumulation throughout the island\cite{hall}. Note also that the signal is about 20 times larger than the signal reported by Jedema~\textit{et al.} in a 1D geometry\cite{jedema02} and comparable with the two-terminal signal in the pillar structures used to study the spin current induced magnetisation reversal\cite{katine}.

\begin{figure}
\includegraphics[width = 0.5\textwidth, height = 0.52\textwidth]{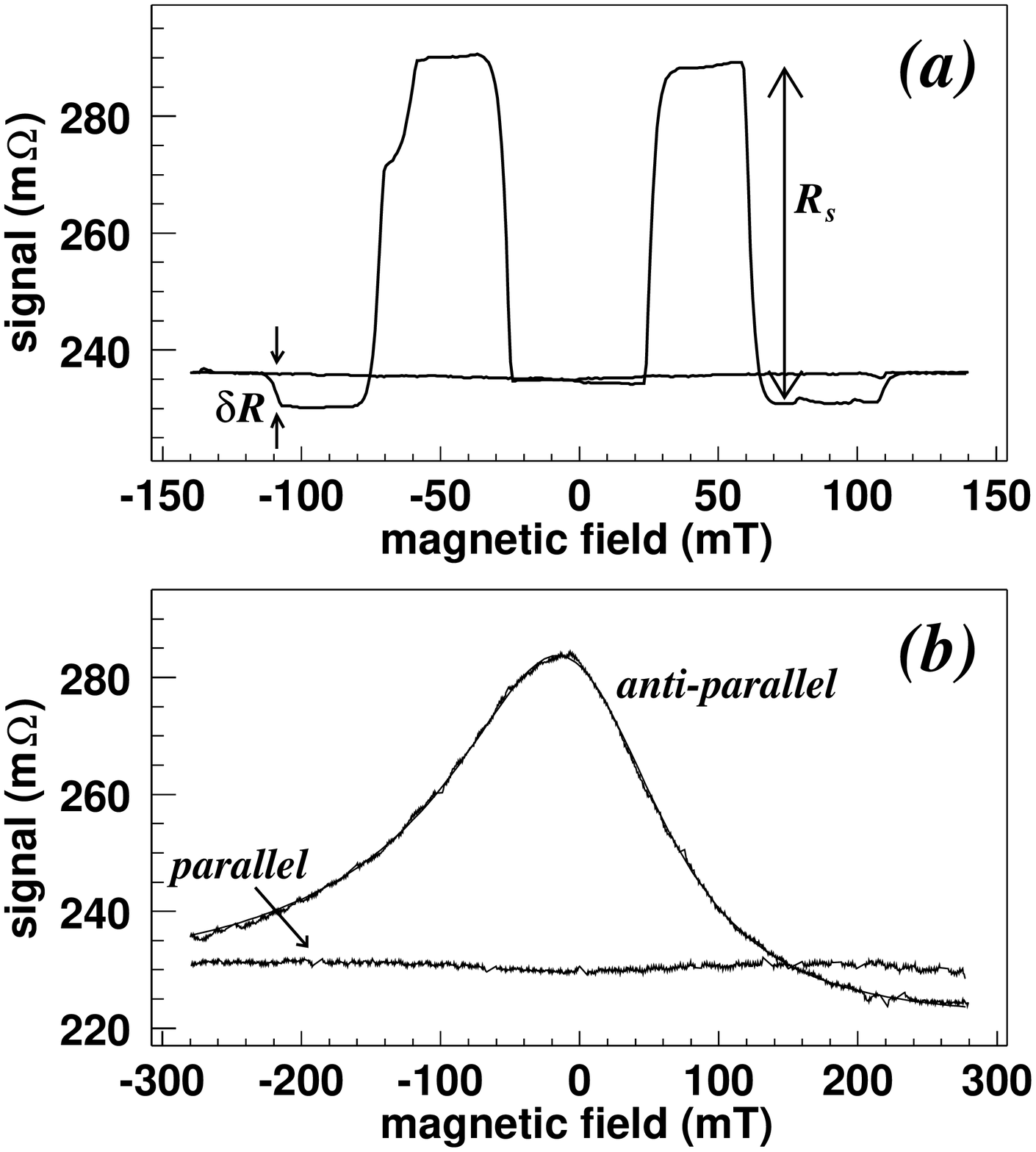}
\begin{picture}(0, 0)
\put(60, 75){\includegraphics[width = 1.0cm, height = 1.35cm]{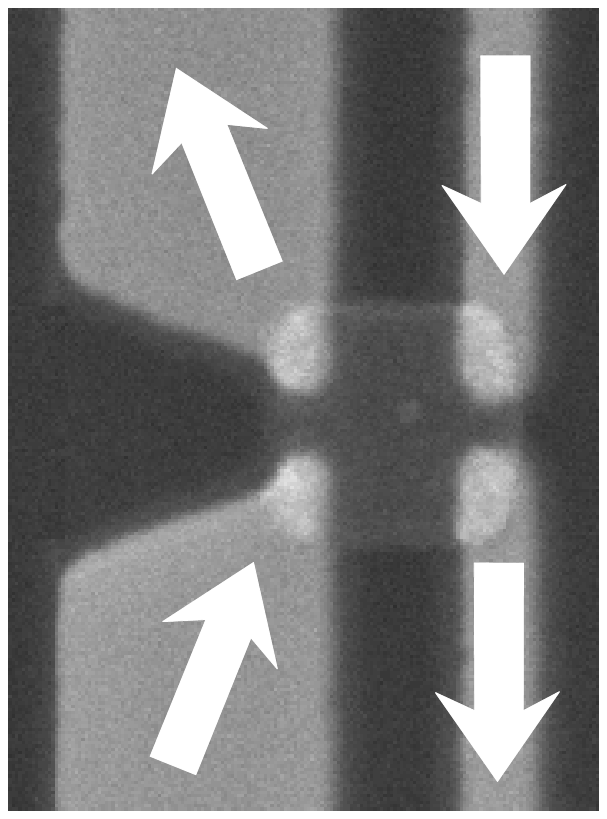}}
\put(-75, 85){\includegraphics[width = 1.0cm, height = 1.35cm]{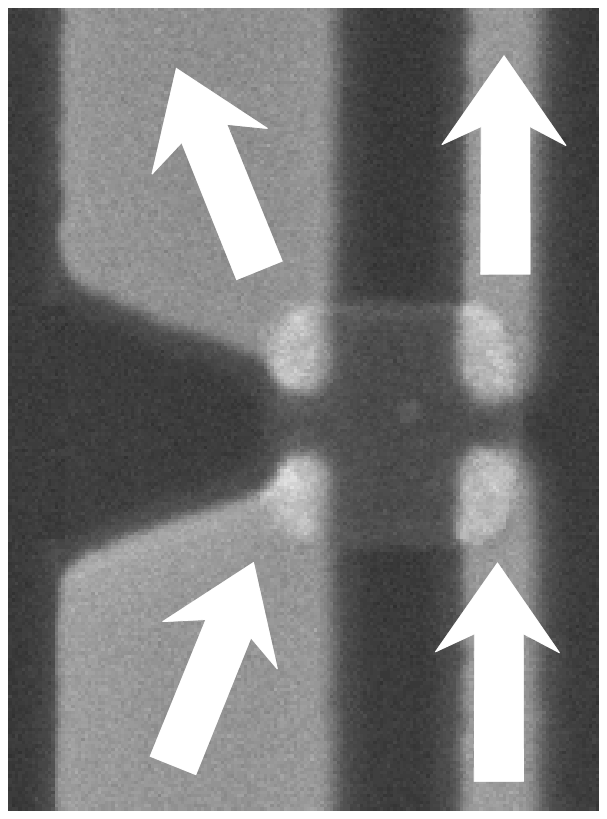}}
\end{picture}
\caption{\label{graph:dipprec} (a) Spin signal at 300~K. The dip that appears at 70~mT is due to the ``anomalous'' magnetic configuration. (b) Spin precession for the same device in the parallel and anti-parallel configurations, in the side configuration. The fitted curve fits the measurement closely. The insets represent the direction of the electrodes' magnetisation in the non-collinear case, i.e. assuming an angle $\phi$ between injector and detector.}
\end{figure}

Fig.~\ref{graph:dipprec}(a) shows a measurement in the ``side'' configuration at room temperature for device B with tunnel barriers of 2~k$\Omega$. Here the spin signal is $R_s = 60$~m$\Omega$. The measurement presents a new feature around 70~mT: while switching from anti-parallel to parallel, the signal dips $\delta R \approx 5$~m$\Omega$ below the signal in the parallel configuration. This can be explained by assuming an ``anomalous'' configuration, see Fig.~\ref{fig:chempot}(c), where only one narrow electrode has reversed. The detected signal is the lowest and equals the ohmic resistance $R_{ohm}$. At a higher field of 120~mT, the other narrow electrode also flips, returning to the parallel configuration and the signal increases by $\delta R = P^2R_{ohm}$. We thus obtain a coarse estimation of $P \approx 16\%$. Another two devices with the same tunnel barrier resistances showed similar behaviour and gave the same spin signal amplitude. A device with higher tunnel barriers showed a room temperature spin signal of 90~m$\Omega$.

To accurately determine the spin relaxation time, we measured the precession of the injected spins under a magnetic field $\mathbf{B}$ applied perpendicular to the device, in the positive $\mathbf{\hat{z}}$ direction. The component of the spins perpendicular to $\mathbf{B}$ precesses with the Larmor angular frequency $\omega_L = g\mu_B |\mathbf{B}|/\hbar$, with $g \approx 2$ for aluminium\cite{johnson85, jedema02}. To derive the spin signal, we use an approach similar to that of Johnson and Jedema, but now derived explicitly for the 0D geometry\cite{hernando}.

Let us assume $\mathbf{m}_{inj}, \mathbf{m}_{det}$ perpendicular to $\mathbf{B}$ and be $\phi$ the angle between them. At $t=0$, spins parallel to $\mathbf{m}_{inj}$ are injected in the island and precesses. The contribution to the detected signal at a later time $t$ is proportional to $\exp(-t/\tau_{sf})\cos(\omega_Lt + \phi)$, where the exponential factor accounts for spin flip scattering. Integrating over the possible times $[0, +\infty)$, the spin accumulation signal becomes:

\begin{equation}
R_s = \frac{P^2\tau_{sf}}{e^2\nu_{DOS}\hat{V}} \frac{\cos\phi - \omega_L\tau_{sf}\sin{\phi}}{1+\omega_L^2\tau_{sf}^2}|\mathbf{m}_{inj}|\ |\mathbf{m}_{det}|
\label{eq:prec} 
\end{equation}

a linear combination of an even and an odd function in the field. This equation reduces to eq.~(\ref{eq:spinaccumulation}) for $\mathbf{B} = 0$. 

In the experiment, we apply first a magnetic field in the $\mathbf{\hat{y}}$ direction to set the leads' magnetic configuration to either parallel or anti-parallel. Then, with the $\mathbf{\hat{y}}$ field switched off, we measure the spin signal as a function of the perpendicular field. The resulting spin precession is shown in Fig.~\ref{graph:dipprec}(b) for device B. In the parallel configuration, only a small dependence of the signal on the $\mathbf{B}$ field is detected. In the anti-parallel case, the spin accumulation reaches a maximum at $-20$~mT and decays asymmetrically. We fit the spin signal with eq.~(\ref{eq:prec}) to which we have added a constant term for the background ohmic resistance. The fitted curve, Fig.~\ref{graph:dipprec}(b), is superimposed on the measurement, and agrees very well with the experimental data: we obtain $\tau_{sf} = 62$~ps at room temperature and $\phi = 0.13\pi$. The latter we believe, reflects the fact that the tips of the larger electrodes have a triangular shape and the end domains are not exactly magnetised along the $\mathbf{\hat{y}}$ direction (see Fig~\ref{graph:dipprec}(b) insets). Assuming that only the wide electrodes magnetisation is rotated by $\phi$, $|\mathbf{m}_{inj}| = |\mathbf{m}_{det}| = 2\cos(\phi/2)$, we find $P = 7\%$. These values agree with the results of Jedema~\textit{et al.}\cite{jedema02}. Note that taking into account the detection efficiency $P$ of the component of the signal, the spin accumulation is $(\mu_\uparrow - \mu_\downarrow)/eI = R_s/P = 60$~m$\Omega/7\% = 850$~m$\Omega$, thus dominating the ohmic resistance. Using the diffusion constant for aluminium at room temperature $D = 5\cdot 10^{-3}$~m$^2/$s, the diffusion time $\tau_{dif\!f} = L^2 / D \approx 30$~ps\cite{diffusion} and the escape time $\tau_{esc} = R_{TB}e^2\nu_{DOS}\hat{V} \approx 10^3\ \tau_{sf}$. This shows that the relation $\tau_{dif\!f} < \tau_{sf} \ll \tau_{esc}$ is satisfied.

In conclusion, we have measured zero-dimensional spin accumulation in a mesoscopic aluminium island at 4.2~K and at room temperature and also coherent spin precession. We have observed three contributions to the total detected signal: the overall spin accumulation (being the largest), the ohmic resistance and the effect of the spin current. From the precession measurements, we have determined the spin relaxation time and the orientation of the magnetic leads. The control of the spin accumulation in the dc regime that we have demonstrated opens the way to the study of the island's magnetisation dynamics with time dependent spin injection in the radio-frequency regime.

\begin{acknowledgments}
We thank Andrei Filip for critically reading the manuscript and we acknowledge Gert ten Brink for technical support. This work was supported by MSC$^{plus}$ and NEDO (Project ``Nano-scale control of magnetoelectronics for device applications'').
\end{acknowledgments}


\begin{thebibliography}{99}
\bibitem{sem} For clarity, we have removed from the micrograph the shadow replicas. These are not in electrical contact to the device and do not affect the system.
\bibitem{reciprocity} The configurations in which current and voltage leads are interchanged and the magnetic fields are reversed, are related to each other, according to the reciprocity theorem for four terminal measurements.
\bibitem{hall} Local Hall effects produced by the fringe field of the magnetic leads can be ruled out, as they would appear as an offset of opposite sign in the two parallel configurations, i.e. with all the leads' magnetisations in the positive and in the negative $\mathbf{\hat{y}}$ direction. Also we estimate the possible Hall effect to be 100 times smaller than the measured signal.
\end{thebibliography}
\end{document}